\documentclass[12pt]{article}
\usepackage{html}
\usepackage{epsf}
\usepackage{graphicx}
\usepackage{amssymb}
\begin{document}
\title{MATTERS OF GRAVITY, The newsletter of the APS Topical Group on 
Gravitation}
\begin{center}
{ \Large {\bf MATTERS OF GRAVITY}}\\ 
\bigskip
\hrule
\medskip
{The newsletter of the Topical Group on Gravitation of the American Physical 
Society}\\
\medskip
{\bf Number 44 \hfill December 2014}
\end{center}
\begin{flushleft}
\tableofcontents
\vfill\eject
\section*{\noindent  Editor\hfill}
David Garfinkle\\
\smallskip
Department of Physics
Oakland University
Rochester, MI 48309\\
Phone: (248) 370-3411\\
Internet: 
\htmladdnormallink{\protect {\tt{garfinkl-at-oakland.edu}}}
{mailto:garfinkl@oakland.edu}\\
WWW: \htmladdnormallink
{\protect {\tt{http://www.oakland.edu/?id=10223\&sid=249\#garfinkle}}}
{http://www.oakland.edu/?id=10223&sid=249\#garfinkle}\\

\section*{\noindent  Associate Editor\hfill}
Greg Comer\\
\smallskip
Department of Physics and Center for Fluids at All Scales,\\
St. Louis University,
St. Louis, MO 63103\\
Phone: (314) 977-8432\\
Internet:
\htmladdnormallink{\protect {\tt{comergl-at-slu.edu}}}
{mailto:comergl@slu.edu}\\
WWW: \htmladdnormallink{\protect {\tt{http://www.slu.edu/colleges/AS/physics/profs/comer.html}}}
{http://www.slu.edu//colleges/AS/physics/profs/comer.html}\\
\bigskip
\hfill ISSN: 1527-3431

\bigskip

DISCLAIMER: The opinions expressed in the articles of this newsletter represent
the views of the authors and are not necessarily the views of APS.
The articles in this newsletter are not peer reviewed.

\begin{rawhtml}
<P>
<BR><HR><P>
\end{rawhtml}
%{\bf \Large Contents:}
\end{flushleft}
\pagebreak
\section*{Editorial}

Matters of Gravity has adopted a new publication schedule: it will appear in December and June.  The 
purpose of this change is so that a preliminary description of the GGR sessions of each upcoming April APS meeting can  be shown to the GGR membership before the deadline for submission of an abstract for the April meeting.
The next newsletter is due June 2015.  This and all subsequent
issues will be available on the web at
\htmladdnormallink 
{\protect {\tt {https://files.oakland.edu/users/garfinkl/web/mog/}}}
{https://files.oakland.edu/users/garfinkl/web/mog/} 
All issues before number {\bf 28} are available at
\htmladdnormallink {\protect {\tt {http://www.phys.lsu.edu/mog}}}
{http://www.phys.lsu.edu/mog}

Any ideas for topics
that should be covered by the newsletter, should be emailed to me, or 
Greg Comer, or
the relevant correspondent.  Any comments/questions/complaints
about the newsletter should be emailed to me.

A hardcopy of the newsletter is distributed free of charge to the
members of the APS Topical Group on Gravitation upon request (the
default distribution form is via the web) to the secretary of the
Topical Group.  It is considered a lack of etiquette to ask me to mail
you hard copies of the newsletter unless you have exhausted all your
resources to get your copy otherwise.

\hfill David Garfinkle 

\bigbreak

\vspace{-0.8cm}
\parskip=0pt
\section*{Correspondents of Matters of Gravity}
\begin{itemize}
\setlength{\itemsep}{-5pt}
\setlength{\parsep}{0pt}
\item Daniel Holz: Relativistic Astrophysics,
\item Bei-Lok Hu: Quantum Cosmology and Related Topics
\item Veronika Hubeny: String Theory
\item Pedro Marronetti: News from NSF
\item Luis Lehner: Numerical Relativity
\item Jim Isenberg: Mathematical Relativity
\item Katherine Freese: Cosmology
\item Lee Smolin: Quantum Gravity
\item Cliff Will: Confrontation of Theory with Experiment
\item Peter Bender: Space Experiments
\item Jens Gundlach: Laboratory Experiments
\item Warren Johnson: Resonant Mass Gravitational Wave Detectors
\item David Shoemaker: LIGO Project
\item Stan Whitcomb: Gravitational Wave detection
\item Peter Saulson and Jorge Pullin: former editors, correspondents at large.
\end{itemize}
\section*{Topical Group in Gravitation (GGR) Authorities}
Chair: Beverly Berger; Chair-Elect: 
Deirdre Shoemaker; Vice-Chair: Laura Cadonati. 
Secretary-Treasurer: Thomas Baumgarte; Past Chair:  Daniel Holz;
Members-at-large:
Curt Cutler, Christian Ott,
Andrea Lommen, Jocelyn Read,
Kimberly Boddy, Steven Drasco,
Sarah Gossan, Tiffany Summerscales.
\parskip=10pt

\vfill
\eject

\vfill\eject

\section*{\centerline
{we hear that \dots}}
\addtocontents{toc}{\protect\medskip}
\addtocontents{toc}{\bf GGR News:}
\addcontentsline{toc}{subsubsection}{
\it we hear that \dots , by David Garfinkle}
\parskip=3pt
\begin{center}
David Garfinkle, Oakland University
\htmladdnormallink{garfinkl-at-oakland.edu}
{mailto:garfinkl@oakland.edu}
\end{center}

Jacob Bekenstein has been awarded the APS Einstein Prize.

Stanley Deser and Charles Misner have been awarded the Einstein Medal of the Albert Einstein Society.

Duncan Brown, Guido Mueller, Maria Alessandra Papa, and Robert Schofield have been elected APS Fellows.

Hearty Congratulations!

\section*{\centerline
{Centenial of General Relativity Speakers Bureau}}
\addtocontents{toc}{\protect\medskip}
\addcontentsline{toc}{subsubsection}{
\it GR Centenial Speakers Bureau, by Deirdre Shoemaker}
\parskip=3pt
\begin{center}
Deirdre Shoemaker, Georgia Institute of Technology
\htmladdnormallink{deirdre-at-gatech.edu}
{mailto:deirdre@gatech.edu}
\end{center}

2015 marks the centennial of Albert Einstein's lectures first describing his theory of general relativity. The American Physical Society Topical Group in Gravitation is organizing the Centennial of General Relativity Speakers Bureau to provide opportunities for experts in astronomy, cosmology, and general relativity to visit colleges, universities, schools, and communities to give public talks, lectures, meet with students and faculty, or participate in other events. Assistance with travel funding for the speaker is available through this program, especially for minority serving institutions and for schools with little or no research activity in physics and astronomy.

For more information and to request a speaker please visit the website:  
\url{http://apsggr.org/?page_id=24}

\section*{\centerline
{GGR $\to$ DGR}}
\addtocontents{toc}{\protect\medskip}
\addcontentsline{toc}{subsubsection}{
{\it GGR} $\to$ \it DGR, by Nicolas Yunes}
\parskip=3pt
\begin{center}
Nicolas Yunes, Montana State University
\htmladdnormallink{nyunes-at-physics.montana.edu}
{mailto:nyunes@physics.montana.edu}
\end{center}

%What is the APS
Many of you probably already know about the American Physical Society (the APS) -- one of the most important physics organizations, representing over 50,000 physicists from all over. The APS sponsors meetings to promote the exchange of new results in physics, the most relevant of which for the gravity community is the April APS Meeting. The APS also publishes some of the most prestigious journals in our field, including the Physical Review D and Physical Review Letters. Moreover, the APS also advocates for physics education and science education in general, and in particular, in front of Congress, promoting funding for physics research at NSF, DOE, and NASA.  

%Breakdown of Units.
What many of you may be less familiar with is the role of ``units'' within the APS. When physicists are left to interact freely, they will natural interact and coalesce to form ``units'', which in the APS can be classified as follows:
\begin{itemize}
\item {\bf{Sections}}: regional groups that foster a spirit of community. 
\item {\bf{Forums}}: national groups built to address broad issues, such as graduate student education or international cooperation. 
\item {\bf{Technical Units}}: national groups that coalesce around a common interest in physics. Technical units are of two types: {\emph{Divisions}} and {\emph{Topical Groups}}, the main difference being primarily one of size -- divisions have a total number of members that exceeds 3\% of total APS membership, while topical groups do not. Their mission, however, is similar: to bring together scientists with overlapping interests in physics and foster cooperation and communication.
\end{itemize}
%What's the difference between divisions and topical groups? 
But size matters. Divisions play a much more important role in the APS governance. They have the right to a permanent seat in the APS council, with greatly enhanced opportunities to advocate for their fields. Divisions also have a higher likelihood of getting plenary speakers for the April Meeting, giving much broader visibility to their field. In addition, divisions receive a larger budget allocation from the APS and offer \$600 in student travel support, compared to the \$300 that Topical Groups offer. A larger budget allocation also means that they usually provide food and wine at their business meetings, during the April APS meeting, which are open to all members. 

%Who belongs to GGR and who does it represent? 
Our community, gravitational physics, is represented by the Topical Group in GRavitation (GGR) in APS. Their members specialize in wide range of disciplines, including analytical and computational studies of general relativity, mathematical and numerical relativity, tests of general relativity and experimental relativity as a whole, modified theories of gravity, relativistic astrophysics, cosmology, quantum gravity, and gravitational wave detection, to name a few. GGR was established in 1995, through an effort spearheaded by Beverly Berger with the invaluable assistance of Stan Whitcomb, Neil Ashby, and others. From the beginning, LIGO has been a key part of GGR, but by no means is GGR only a LIGO community. Both GGR and the LSC owe their existence to the growth of interest in gravitation which the past two decades have seen. 

%Why does GGR matter? 
GGR provides a broadly-based but external voice to allow the gravitational physics community to advocate for its interests and, as a member of GGR, you benefit directly from the activities the GGR executive committee organizes. GGR organizes a number of sessions in the APS April meeting, including talks on quantum gravity, cosmology, relativistic astrophysics, and gravitational waves. This meeting, in fact, has become one of our main forums for releasing new results. GGR sponsors and selects the winner of the APS Einstein Prize, awarded to Jacob Bekenstein (2015), Irwin Shapiro (2013), Ted Newman (2011), Jim Hartle (2009), Rai Weiss and Ron Drever (2007), Bryce DeWitt (2005), and John Wheeler and Peter Bergmann (2003). GGR has named over sixty APS fellows, an important and distinct honor signifying recognition by one's professional peers in the physics community. And, of course, GGR provides the much needed student travel support in these days of difficult funding.

%News of Division status. 
These activities don't just happen -- it is the GGR executive officers and the support for GGR membership that make them happen. Growth in GGR membership is one of the best arguments we have to leverage more benefits for all of us, including increased funding in physics and gravitation and more visibility at conferences. Joining GGR was your way of standing up to be counted in the community of gravitational physicists and you have done just that. GGR is the largest topical group in the APS, but now it has grown and matured to the point that we will be able to petition APS to become a Division. Your membership did that. 

%But it's not over yet. 
But we are not there yet. To become a division, we must maintain our membership numbers above 3\% of total APS membership for two consecutive years. (at this time we are either just slightly below or just slightly above the 3\% threshold). I have no doubt that this can be achieved, if we all work together to retain old members and make new ones. Thank you once more for your continuous support. Achieving this milestone makes me immensely proud to be a member of this community. 

Nicolas Yunes
Membership Czar 2012-2014

\section*{\centerline
{GGR program at the APS meeting in Baltimore, MD}}
\addtocontents{toc}{\protect\medskip}
\addcontentsline{toc}{subsubsection}{
\it GGR program at the APS meeting in Baltimore, MD, by David Garfinkle}
\parskip=3pt
\begin{center}
David Garfinkle, Oakland University
\htmladdnormallink{garfinkl-at-oakland.edu}
{mailto:garfinkl@oakland.edu}
\end{center}

We have a very exicting GGR related program at the upcoming APS April meeting in Baltimore, MD, in 2015 the Centenial of General Relativity.  Our Chair-Elect, Deirdre Shoemaker, did 
an excellent job of putting together this program.  
\vskip0.25truein
{\bf Note that the deadline for submitting an abstract for this meeting is Friday, January 9, 2015 at 5:00 pm EST}
\vskip0.10truein
{\centerline {abstracts can be submitted at}}
\url{http://www.aps.org/meetings/april/scientific/index.cfm}
\vskip0.25truein
At the APS meeting three of the plenary talks will be devoted to gravity: 
\vskip0.2truein
Clifford Will: Precision Tests of the Theory of General Relativity\\
Stuart Shapiro: Sources and Detection of Gravitational Waves\\
James Hartle: Quantum Gravity and Cosmology\\

There will be several invited sessions of talks sponsored by the Topical Group in Gravitation as follows:\\

Innovative Computing in Relativity\\
(Peter Diener, Zachariah Etienne, Tyson Littenberg)\\
with DCOMP\\
 
Neutron Stars as Laboratories for Neutrino, Nuclear and Gravitational Physics\\
(Benjamin Lackey, Evan O’Connor, Jorge Piekarewicz)\\
with DNP\\

Detecting GWs from the Ground and in Space\\
(Shane Larson, Jason Hogan, David Shoemaker)\\
with DAP\\

Astrophysical Black holes on all mass scales\\
with DAP\\

Quantum Gravity in the 100th Anniversary of General Relativity\\
(Joseph Polchinski, Thomas Faulkner, Walter Goldberger)\\
with DPF\\

100 Years of GR, 20 Years of GGR Looking back and looking forward (panel session)\\
(Rainer Weiss, Gabriela Gonzalez, James Hartle and Jorge Pullin)\\

Precision experimental measurements of gravitation\\
(Michael Hohensee, JamesFaller, Charles Hagedorn)\\
with GPMFC\\

History of Relativity\\
(Saul Teukolsky, Diana Kormos-Buchwald )\\
with FHP\\

GGR Prize Session\\

\vfill\eject
\section*{\centerline
{New Frontiers in Dynamical Gravity}}
\addtocontents{toc}{\protect\medskip}
\addtocontents{toc}{\bf Conference reports:}
\addcontentsline{toc}{subsubsection}{
\it New Frontiers in Dynamical Gravity, 
by Helvi Witek}
\parskip=3pt
\begin{center}
Helvi Witek, DAMTP, University of Cambridge
\htmladdnormallink{h.witek-at-damtp.cam.ac.uk}
{mailto:h.witek@damtp.cam.ac.uk}
\end{center}

In March 2014 we hosted the workshop ``New frontiers in dynamical gravity'' or, in short, 
``Gauge/gravity duality 
meets Numerical Relativity meets fundamental math'', 
at the futuristic site of DAMTP at the University of Cambridge,
organised by P.~Figueras, H.~Reall, U.~Sperhake and myself.

The gauge/gravity correspondence provides a powerful tool to understand strongly coupled 
conformal field theories in $D-1$ dimensions  
by investigating gravity in asymptotically anti-de Sitter (AdS) spacetimes in $D$ dimensions and vice versa.
Nowadays, the duality is available in many different flavours,
employing calculations in GR to explore vastly different fields such as hydrodynamics and
condensed matter physics.
On the gravity side, which was the main focus of our workshop,
this requires finding solutions in AdS and understanding their properties and stability
as well as their dynamical evolution in time.
Many of these issues cannot be tackled by ``pen and paper'' calculations and demand a numerical treatment using both
``soft'' and ``hard'' numerics~\footnote{These labels have been introduced by R. Emparan 
at yet another fantastic workshop on ``Numerical Relativity and High Energy Physics'' held at Madeira in 2011 to distinguish
between solving ODEs or simplified PDEs using computer algebra tools like Mathematica on a desktop and 
the more complex computations of BH dynamics 
(in at least 2+1 dimensions) 
requiring typically hundreds of computer cores.}.

In this workshop we brought together leading experts in these fields.
The schedule of the conference -- typically two one-hour long overview talks in the morning
and four half-hour talks in the afternoon -- 
left plenty of time for fruitful discussions, the exchange of ideas and the launching of new collaborations.
The slides as well as the group photo are available on the conference's website
\url{http://www.ctc.cam.ac.uk/activities/adsgrav2014/}.
%

%%%%%%%%%%%%%%%%%%%%%%%%%%%%%%%%%%%%%%%%%%%%%%%%%%%%%%%%%%%%%%%%%%%%%%%%%%%%%%
%\section{Summary of talks}
%%%%%%%%%%%%%%%%%%%%%%%%%%%%%%%%%%%%%%%%%%%%%%%%%%%%%%%%%%%%%%%%%%%%%%%%%%%%%%
While the main focus of our workshop was gravity in AdS, there are still many open questions concerning
the stability of black holes (BHs) even in four dimensional, asymptotically flat spacetimes.
This topic has been in the spotlight of a number of talks,
kicking off with M.~Dafermos who discussed the nature of BH singularities -- we have
learned that the generic BH singularity
might not be space-like after all -- and its importance for the
(strong) cosmic censorship conjecture.
This is closely related to still open questions about the linear and non-linear stability of BHs
which is an active field of research.
In his talk, S.~Hollands discussed the thermodynamic stability of black objects 
(in four and higher dimensional spacetimes)
and its implications for their dynamical properties.
Although the ``standard'' lore states that BHs in four-dimensional, asymptotically flat spacetimes are stable
this, in fact, only refers to mode stability which excludes a vast number of possibly growing solutions.
G.~Holzegel presented a very pedagogical summary of the state-of-the-art
of the mathematical understanding of the 
stability of BHs: while it has been proven that Schwarzschild BHs are linearly stable, he reminded us 
that the linear stability in a strict mathematical sense even of the Kerr BH is still a completely open question.
Indeed, it has recently been shown that extremal BHs do suffer from an instability, 
which failed to show up in a mode analysis.

Nevertheless, such a mode analysis giving the characteristic response of a BH towards perturbations
provides substantial insight into BH phenomenology.
C.~Warnick discussed the construction of quasi-normal modes (QNMs) for AdS BHs.
As an illustrative example he presented computations of excitations of the Schwarzschild-AdS BH induced by scalar perturbations.
Despite the importance and beauty of QNMs 
they do not form a complete basis for AdS BH spacetimes, thus hinting at more physics that
await to be uncovered.

Another novel result which may make us revise our understanding of
classical BHs was presented by C. Herdeiro
in the form of hairy rotating BH solutions. Inspired by recent studies 
on the superradiant instability of Kerr BHs in the presence of massive fields
they constructed a family of rotating solutions with (complex) scalar hair 
which branches off at the onset of the superradiant instability and can evade the ``no-hair'' theorem.
The specific properties of these solutions such as their stability are still under investigation.

D. Hilditch discussed new results concerning critical phenomena in asymptotically flat spacetimes.
These phenomena, marking a phase-transition between BH formation versus dispersion
during scalar field collapse, are well understood in spherical symmetry.
However, the axisymmetric case still lacks a complete understanding. In the literature one can find 
various, sometimes contradictary results, which have been revisited with 
the new spectral (Numerical Relativity) code {\textsc{BAM-PS}}. 
Ongoing, challenging numerical simulations probe the phase-space near the onset of criticality.
We are looking forward to hear more about the final results.

A couple of years ago investigations of critical phenomena in spherically symmetric 
AdS spacetimes brought forward a surprising outcome,
which has been summarized in P. Biz\'{o}n's talk.
While the asympotically flat case provided a clear-cut transition between
collapse and dispersion this picture changes dramatically in AdS. 
If the scalar field fails to form a BH upon its first interaction, 
in subsequent reflections at the AdS (timelike) boundary the field is focused more and more,
thus eventually yielding collapse.
Given that this mechanism generically involves a cascade towards higher energies 
and a transition from ``pure'' AdS to an AdS BH spacetime 
this phenomenon has been termed ``turbulent instability''.
However, the understanding of this instability is still in its infancy. 
Although pure AdS appears to be generically non-linearly unstable 
there do exist ``islands of stability'' -- a fine-tuned set in the parameter space of perturbations
which evade this fate. 
Both M. Maliborski and S. Liebling gave us an update on their ongoing work 
attempting to reveal the underlying mechanisms.

These outcomes inspired investigations of gravitational turbulence in 
AdS BH spacetimes as L. Lehner demonstrated in his talk. Indeed, it has been 
observed in perturbative calculations 
that the interplay between the AdS BH and perturbations provokes a non-linear mode-coupling and 
cascade towards higher energies in certain regimes
defined by the gravitational analog of the Reynolds number.
%Whether gravitational turbulence requires a confining mechanism as is provided for example by the AdS boundary
%or whether it can also appear in asymptotically flat spacetimes is not yet completely understood.
Additionally, these studies motivated numerical simulations re-visiting turbulence in hydrodynamics.
Features such as the formation and annihilation of vortices have been illustrated in 
beautiful animations.

A further application of the fluid/gravity duality has been discussed in a series of talks by
by R. Janik, M. Heller, P. Romatschke and K. Balasubramanian.
They focused on shock-wave collisions in AdS spacetimes which 
can be interpreted as interactions between a strongly coupled plasma
and have specifically been employed to model heavy-ion collisions at RHIC or LHC.
Before starting to attack interesting physics' questions one has to solve a number
of technical (numerical) challenges including
(i) the representation of AdS spacetimes (i.e. Poincar\'{e} versus global AdS), 
(ii) the specific space+time formulation of Einstein's equations, which needs to be a well-posed
initial boundary value problem,
and (iii) means to extract relevant physical information from the boundary stress-energy tensor.
These challenges and novelties for Numerical Relativity have recently been tackled.
Simulations of the collisions of shockwaves in (2+1) AdS indeed allowed to represent
the strong dynamics of the collisions of heavy ions in the field theory counterpart. 
It is exciting to note, that a proposed model using 
both GR simulations for the dynamical collision part followed by a hydrodynamic description 
after the equilibration
shows excellent agreement with real-world particle collision. 
For example, particle spectra resulting from Pb-Pb collisions at the LHC
have been fitted well with this new model
while pure hydrodynamical models have been off by an order of magnitude.

T. Wiseman reported on the latest news concerning plasma flows with space-dependent temperature profiles.
On the gravity side these fluids can be represented by recently discovered AdS BHs with a non-Killing horizon.
He gave a predagogical summary of the numerical methods to construct these stationary solutions 
which requires solving a set of PDEs with  mixed elliptic and hyperbolic characteristics.
Via the correspondence, information about the surface gravity and (linear) velocity of the horizon
of these BHs provides insight into physical parameters of the flowing plasma such as its (local) velocity.
For a certain region of the phase-space, this plasma velocity seems to diverge
hinting at an instability, possibly related to turbulent behaviour.

A further exciting application of the gauge/gravity duality represents itself in
the correspondence between gravity and condensed matter physics.
In particular, J. Gauntlett, O. Dias and J. Santos discussed the holographic duals 
of metals, insulators and superconductors.
A complete understanding of their phase transitions in condensed matter physics is still lacking.
Via the correspondence these materials can be modelled in the context of gravity in AdS
coupled to a Maxwell and, possibly, scalar fields.
Using a periodic potential or constructing AdS BHs with non-Killing, ``floppy'' horizons 
allow to mimic so-called Q-lattices with broken translational invariance.
Computations on the gravity side recover, e.g., metal/insulator transitions and 
the (DC) conductivity known from the solid-state physics side.
Moreover, recent investigations predict the existence of new insulator and metal phases and it
will be interesting to see whether they can be discovered in real materials.

A.~Ishibashi addressed the instability of AdS and its connection with singularity theorems in AdS
by studying %(the symmetry and momentum of) 
Bianchi black branes.
The key requirements of known rigidity theorems (in asymptotically flat spacetimes)
include the weak energy condition and the compactness and analyticity of BH horizons.
While the model presented in this talk still satisfies the weak energy condition it is possible to construct
solutions which have either non-smooth or non-compact horizons, such as stationary solutions with 
non-Killing horizons, and thus evade the rigidity theorem.

So far we have seen a number of applications of the gauge/gravity duality which exploit well-understood physics
on the gravity side to learn about hard-to-tackle problems in the dual field theory.
However, we can also employ the duality to understand fundamental issues in 
quantum gravity which should emerge near a BH singularity.
Bearing in mind that bulk BHs can be described as thermal states in the boundary field theory 
this opens up the exciting possibility to learn about quantum gravity in AdS by investigating the gauge theory.
G.~Horowitz presented a holographic model which employs a conformal field theory,
specifically $\mathcal{N}=4$ super Yang-Mills theory 
in an anisotropic generalization of de Sitter spacetimes,
to explore the nature of (quantum) gravity near singularities in the bulk theory. 
V. Hubeny discussed how extremal surfaces and geodesics in Vaidya-AdS spacetimes, 
modelling BH formation through gravitational collapse, 
can be used to probe the spacetime region close to the singularity.
In contrast to static spacetimes they can penetrate the (event) horizon in the dynamical setting
at hand and, furthermore, exhibit a strikingly rich structure.
The properties of these geodesics and surfaces can be interpreted in terms of
thermodynamic quantities such as entanglement entropy.
In a related talk M. Taylor described how holography might help to better understand
the nature of the BH interior. One of the key concerns are related to our 
still poor comprehension of the horizon and the information loss paradox,
driven by (semi-) classical pictures.
She gave a short review on recent proposals such as firewalls and fuzzballs 
to resolve these issues,
and focused on a solution using BH microstates.
These states represent horizonless ``stringy'' geometries, thus
escaping the ``teething troubles'' of our classical models.
Through a coarse-graining over these geometries we can recover the familiar classical BH picture.
C. Pope gave a talk reviewing the thermodynamics of BHs in asymptotically AdS spacetimes.
M. Rangamani presented his ongoing work on computing the entanglement entropy in a boundary field theory
using minimal surfaces in the bulk. While this construction in principle is well understood,
ensuring the causality of these solutions is fundamental.
In fact, causality can be violated in the presence of time-like singularities in the bulk.
He discussed a number of examples including 
negative-mass Schwarzschild-AdS solutions or charged scalar solitons with positive boundary energy.
The requirement that causality should hold therefore yields non-trivial constraints on these extremal surfaces.

J. Armas illustrated how membranes in hydrodynamics can be mimicked using blackfolds. 
In particular, he discussed how this approach in higher order perturbation theory can be employed 
to compute transport coefficients for surfaces or (mem-) branes in hydrodynamics.
K. Skenderis discussed the dynamics of non-equilibrium solutions using the fluid/gravity duality. In particular,
it is possible to capture the dynamics in the long wavelength regime, when the field theory is close
to a thermal equilibrium, using a hydrodynamic description. This regime can be modelled in the gravity dual
by constructing solutions in a gradient expansion giving a good prescription at long distances
and late times.
One such solution is the Robinson-Trautman metric, which can be seen as the dynamical (non-linear)
version of algebraically special perturbations of the Schwarzschild-AdS solution,
allowing us to investigate the effect of non-linearities and the approach to equilibrium.

A further focus of this meeting were the properties of BHs in higher dimensional, 
but asymptotically flat spacetimes.
R. Emparan gave a pedagogical review about his work on BHs in the large-D limit. 
In a nutshell, one can treat the spacetime dimension $D$ simply as a further parameter. 
Then, in the limit that the dimension becomes very large the equations of motion 
decouple into a far region (essentially described by flat spacetime), 
and a near region, in which the gravitional field is concentrated in a thin shell around the BH.
Employing this approach facilitates an analytic treatment of generically rather complicated 
problems, such as the computation of QNMs or the understanding of the 
stability of higher-dimensional BHs.

In order to study the non-linear stability properties of singly-spinning Myers-Perry solutions 
in finite spacetime dimensions a fully numerical treatment is mandatory.
M. Shibata presented new results of evolutions of Myers-Perry BHs in five dimensions
obtained with the improved {\textsc{Sacra-5D}} code which uses the constraint damping mechanism
facilitated in the so-called Z4c formulation of Einstein's equations. 
These new numerical simulations are in excellent agreement 
with a perturbative calculation published earlier this year.
Because this type of ``hard'' numerical evolutions are extremely demanding in terms of computational resources,
they have not yet covered the entire parameter space and we are looking forward to 
read and hear about the latest results.

%%%%%%%%%%%%%%%%%%%%%%%%%%%%%%%%%%%%%%%%%%%%%%%%%%%%%%%%%%%%%%%%%%%%%%%%%%%%%%
%\section{Announcements/Social events}
%%%%%%%%%%%%%%%%%%%%%%%%%%%%%%%%%%%%%%%%%%%%%%%%%%%%%%%%%%%%%%%%%%%%%%%%%%%%%%
Wednesday night we all gathered for a feast at the beautiful Trinity college.
After indulging in excellent food and wine H.~Reall gave an inspiring speech recapturing the first days of our conference
and recalled some amazing anecdotes about I.~Newton. Apparently, once upon a time science could come with 
a large amount of suffering and physical pain.

One of the highlights of this conference was the visit to the {\textsc{Cosmos}} supercomputer
which is part of the DiRAC HPC Facility funded by STFC and BIS. Thank you for the tour, Juha and James!

On Thursday we had a special lunch to celebrate the recent creation by the University of Cambridge of a 
Stephen~W.~Hawking Professorship in Cosmology. This chair was endowed by a donation from Dennis~Avery and
Sally~Wong~Avery. Sadly, Dennis died in 2012 but we were delighted that Mrs~Avery and members of her family
were able to join the workshop participants for this celebration.

G.~Horowitz announced that J.~Santos has been awarded the 2014 General Relativity and Gravitation Young
Scientist Prize by the International Union of Pure and Applied Physics. Congratulations, Jorge!
%

%%%%%%%%%%%%%%%%%%%%%%%%%%%%%%%%%%%%%%%%%%%%%%%%%%%%%%%%%%%%%%%%%%%%%%%%%%%%%%
%\section{Summary}
%%%%%%%%%%%%%%%%%%%%%%%%%%%%%%%%%%%%%%%%%%%%%%%%%%%%%%%%%%%%%%%%%%%%%%%%%%%%%%
\noindent{\bf{Acknowledgements:}}
%%%%%%%%%%%%%%%%%%%%%%%%%%%%%%%%%%%%%%%%%%%%%%%%%%%%%%%%%%%%%%%%%%%%
%The workshop ``New frontiers in dynamical gravity'' has been organised by
%P. Figueras, H. Reall, U. Sperhake and H. Witek.
%
We thank all the participants for their invalueable input in many illuminating talks and discussions
making this workshop a great success.

We acknowledge financial support for this conference from 
the Institute of Physics/ Gravitational Physics Group, STFC, 
the {\it ERC-2011-StG 279363--HiDGR} ERC Starting Grant and 
Intel through the Centre for Theoretical Cosmology in Cambridge.

\vfill\eject
\section*{\centerline
{Frontiers of Neutron Star Astrophysics}}
\addtocontents{toc}{\protect\medskip}
\addcontentsline{toc}{subsubsection}{
\it Frontiers of Neutron Star Astrophysics, 
by David Nichols}
\parskip=3pt
\begin{center}
David Nichols, Cornell University 
\htmladdnormallink{david.nichols-at-cornell.edu}
{mailto:david.nichols@cornell.edu}
\end{center}

On May 29 and 30, 2014, a meeting called ``Frontiers of Neutron Star 
Astrophysics'' was held at Cornell University to review open problems in 
neutron-star astrophysics and discuss future directions for their
solution.
It was also an opportunity to celebrate the 65$^{\rm th}$ birthday of Jim 
Cordes and the 60$^{\rm th}$ birthday of Ira Wasserman at a conference
banquet on the evening of the 29$^{\rm th}$.
Over the two days, a wide range of topics were covered in seventeen invited
talks, eleven contributed talks, and a concluding panel discussion.
These areas include mechanisms of supernova explosions; properties of radio
pulsars, magnetars, and accreting neutron stars; evolution of magnetic
fields; models of neutron star interiors and equation of state; binary
neutron-star mergers and their accompanying gravitational waves and 
electromagnetic counterparts; and tests of general relativity using pulsars.
Many of the slides from the talks are available on the conference website
at \url{http://www.astro.cornell.edu/nsfrontiers/}.
The conference would not have been possible without the work of the chair
of the scientific committee (Lars Bildsten) and its members (Phil Arras,
David Chernoff, Jim Cordes, \'Eanna Flanagan, Dong Lai, Saul Teukolsky, 
and Ira Wasserman), and support from the Cornell Department of Astronomy 
and Center for Radiophysics and Space Research.

The conference started off with a bang with Adam Burrows' talk on the status
of simulations of core collapse supernovae and the mechanisms that drive the
collapse.
Lars Bildsten's presentation followed along a similar note by discussing
the link between superluminous supernovae and the birth of magnetars.
This first session concluded with a lecture by Duncan Lorimer on the 
populations of radio pulsars, focusing on how the current catalog of known
pulsars can help understand the formation and evolution of neutron stars in
a variety of environments.

Vicki Kaspi began the next session with a talk entitled 
``Magnetars and their ilk.'' 
David Kaplan described the properties of nearby neutron stars that are
emitting thermally, with an emphasis on the optical and x-ray emission from
isolated neutron stars and what this emission can reveal about their physics.
Andrew Cumming then gave the last talk of the section with a presentation on
the evolution of magnetic fields in the crust of neutron stars.

The third session of the day began with talks from the birthday honorees.
Jim Cordes discussed the detection of fast radio bursts and commented on
whether these are likely to be local or cosmological sources.
Ira Wasserman elaborated on the effects of superconductivity on the 
behavior of neutron-star magnetic fields and how this could affect 
pulsar-timing measurements.
The third presentation was given by Armen Sedrakian who described 
superfluidity and pair-breaking processes in baryonic matter and exotic 
cooling mechanisms in neutron stars.
The final talk of the session was by Andrzej Szary, and he presented
a model of a partially screened gap in pulsars that helped to explain 
properties of the radio and x-ray emission.

The last group of talks on the first day commenced with Anatoly Spitkovsky
on the computational modeling of pulsar magnetospheres.
Next was Konstantinos Gourgouliatos who commented on the role of Hall drift
in determining the braking indices of young pulsars.
Describing a model for radio emission from magnetars was George Melikidze,
and ending the session was Wojciech Lewandowski, who discussed the emission
from gigahertz-peaked pulsar spectra.

Beginning the second day was Andrei Beloborodov, who reviewed the mechanisms
responsible for the activity in pulsars and magnetars.
David Tsang presented a mechanism by which tidal gravitational fields in a
compact binary containing a neutron star could resonantly shatter the star's
crust leading to a precursor to short gamma-ray bursts.
Next, Scott Ransom gave a survey of several accurately timed pulsars in 
binaries (and hierarchical triples) that allow for measurements of the 
systems' post-Keplerian parameters and mass-radius relationship.
The session concluded with Sebastien Guillot describing how measurements of
the spectra of low-mass x-ray binaries can be modeled to determine the 
mass-radius relationship, and the neutron stars' equation of state, 
accordingly.

Michael Kramer started the next section of talks, in which he reviewed the
ability of the most relativistic binary pulsars to test the predictions of
general relativity in the strong-field regime and to put constraints on 
modified gravitational theories.
Following this talk, Maura McLaughlin reviewed the prospects for using 
a pulsar-timing array to detect gravitational waves from several types of 
low-frequency sources.
The last speaker of the session was Anna Watts, who described recent progress
in understanding x-ray bursts and burst oscillations in neutron stars.

The penultimate session began with a talk by Edo Berger explaining the 
observations supporting compact-object mergers being the progenitors
of short gamma-ray bursts.
Ben Lackey then discussed the prospects for measuring information about
neutron-star equation of state from the gravitational waves from binary
neutron-star mergers in interferometric gravitational-wave detectors.
James Clark described how burst gravitational-wave searches might be able
to detect gravitational waves emitted after the merger of binary neutron
stars, when the merger forms a hypermassive neutron-star.
Finally, Brian Metzger gave the last talk in this part on kilonovae, 
powered by the decay of r-process elements, being an important 
electromagnetic counterpart to the gravitational-wave signal from binary
neutron-star mergers.

In the final session of talks, Francois Foucart discussed the status of
numerical-relativity simulations of black-hole--neutron-star binaries and
binary neutron-star mergers.
Marc Favata described the systematic errors that arise when incomplete 
gravitational-wave models are used to measure the neutron-star equation of
state from binary neutron-star mergers.
The last talk of the conference was delivered by Sinead Walsh, who reviewed
the status of gravitational-wave searches for unknown isolated neutron stars.

Before the conference ended, several of the invited speakers served on a 
panel discussion about where the field of neutron-star astrophysics will be
headed in the next decade.
While there were a range of opinions about the specific results that would be
found, the general consensus was that the prospect for new discoveries is
good.

\vfill\eject
\section*{\centerline
{Quantum Information in Quantum Gravity}}
\addtocontents{toc}{\protect\medskip}
\addcontentsline{toc}{subsubsection}{
\it Quantum Information in Quantum Gravity, 
by Mark Van Raamsdonk}
\parskip=3pt
\begin{center}
Mark Van Raamsdonk, University of British Columbia 
\htmladdnormallink{mav-at-phas.ubc.ca}
{mailto:mav@phas.ubc.ca}
\end{center}

  During recent years, a truly remarkable connection between the physics
of spacetime/  gravitation and the physics of quantum information has
emerged, largely via the AdS/CFT correspondence in string theory.
While surely still far from being understood completely, there is now
intriguing evidence that the structure and geometry of spacetime in
these examples is related directly and quantitatively to the structure
of entanglement of the fundamental degrees of freedom of the
theory. Further, even the dynamics of spacetime, at least in the limit
of weak curvature, has been understood to emerge from fundamental
constraints obeyed by entanglement. The new ideas have presented
challenges to some long-held beliefs about gravitational physics,
famously including the smoothness of spacetime at black hole horizons.
In order to present and discuss the latest work on these exciting
developments, the conference “Quantum Information in Quantum Gravity”
was held during the week of August 18-22, 2014 in Vancouver, Canada.

     The setting for much of the recent work presented at the conference
is the AdS/CFT correspondence, by which the states of certain
conformal quantum field theories are in one-to-one correspondence
with the states of some corresponding quantum theory of gravity.
Several of the talks related to a conjecture by Ryu and Takayanagi
(and its covariant generalization by Hubeny, Rangamani, and
Takayanagi), which suggests that the entanglement entropy (a measure
of quantum entanglement) for some spatial subset of degrees of
freedom in the field theory is directly proportional to the area of a
certain surface in the corresponding spacetime geometry. Specific
topics included: understanding how this proposal can be used to
extract dual spacetime geometry from a CFT state (Myers, Sully),
understanding how the proposal generalizes to include quantum and
higher curvature corrections (Dong, Wall), understanding the quantum
information-theoretic interpretation of more general geometrical
observables in the gravity theory (Hayden), and deriving the proposal
for the case of 2D CFTs (Hartman) with special properties.

      Several of the other talks (Karch, Mathur, Berkooz) concerned a
more general proposal for relating entanglement and geometry, which
suggests that entangling non-interacting subsets of the fundamental
degrees of freedom, amounts to creating a wormhole in spacetime
between two distant (or disconnected) parts of spacetime. This
proposal has recently been dubbed “ER=EPR” by Maldacena and
Susskind.

     Connections between entanglement entropy and gravitational physics
provide a generalization and refinement of the now famous connections
between gravitational quantities and thermodynamic quantities; in
particular the entropy-area connection for black holes. Entanglement
entropy has the property that it can be evaluated for any quantum
state without assumptions about equilibrium. Thus, certain previous
conjectures involving entropy in the gravitational context can be
made more general and precise if the entropy is interpreted as
entanglement entropy (either directly in the gravitational theory, or
for some dual degrees of freedom). Talks by Bousso (on a proof of his
covariant entropy bound in certain contexts) and Marolf (on a version
of the generalized second law) related to these entanglement entropy
generalizations of conjectures about gravitational thermodynamics.

     A subject of great debate in the recent quantum gravity literature is
the “firewall paradox” of Almheri, Marolf, Polchinski, and Sully.
These authors have argued that the maximal entanglement of an old
black hole with its Hawking radiation forbids the local entanglement
of quantum fields across the black hole horizon that would be
necessary to ensure a smooth spacetime there. Thus, according to the
argument, such a black hole must have a singularity (or “firewall”)
at its would-be horizon. Talks by Giddings, Harlow, Verlinde and
Silverstein dealt with various aspects of this and the closely
related black hole information paradox.

        The conference also featured talks aimed at better understanding
entanglement in a purely field theory context, including understanding
entanglement entropies in gapped theories (Nishioka), and understanding
the evolution of entanglement/density matrices corresponding to a long
wavelength subset of degrees of freedom in field theory (Lawrence). In
discrete field theory systems, there is a useful representation of
quantum states that makes the spatial entanglement structure manifest.
This is known as the Multiscale Entanglement Renormalization Ansatz or
MERA. It was conjectured by Brian Swingle that the MERA representation of
a quantum state for a field theory system with gravity dual may be
directly related to how the dual spacetime is encoded. A number of talks
(Swingle, Takayanagi) presented new results relating to this MERA
description of field theory states.

        Overall, the activities of the conference reinforced the impression that
connections between gravity and quantum information represents an
extremely interesting frontier in gravitational research.

Note: slides for many of the talks can be found at:

\url{http://www.maths.dur.ac.uk/~dma0mr/qiqg-ubc/programme.html}

\end{document}